# Coexistence of full-gap superconductivity and pseudogap in two-dimensional fullerides

Ming-Qiang Ren[1], Sha Han[1], Shu-Ze Wang[1], Jia-Qi Fan[1], Can-Li Song[1,2, †], Xu-Cun Ma[1,2, †],

Qi-Kun Xue[1,2,3†]

[1] *State Key Laboratory of Low-Dimensional Quantum Physics, Department of Physics, Tsinghua University, Beijing 100084, China*

[2] *Frontier Science Center for Quantum Information, Beijing 100084, China*

[3] *Beijing Academy of Quantum Information Sciences, Beijing 100193, China*

**Alkali-fulleride superconductors with a maximum critical temperature $T_c \sim$ 40 K exhibit similar electronic phase diagram with unconventional high-$T_c$ superconductors[1-5] where the superconductivity resides proximate to a magnetic Mott-insulating state[3-6]. However, distinct from cuprate compounds, which superconduct through two-dimensional (2D) $CuO_2$ planes[7], alkali fullerides are attributed to the three-dimensional (3D) members of high-$T_c$ family[8,9]. Here, we employ scanning tunneling microscopy (STM) to show that trilayer $K_3C_{60}$ displays fully gapped strong coupling *s*-wave superconductivity that coexists spatially with a cuprate-like pseudogap state above $T_c \sim$ 22 K and within vortices. A precise control of electronic correlations and doping reveals that superconductivity occurs near a superconductor-Mott insulator transition (SMIT) and reaches maximum at half-filling. The *s*-wave symmetry retains over the entire phase diagram, which, in conjunction with an abrupt decline of superconductivity below half-filling, demonstrates that alkali fullerides are predominantly phonon-mediated superconductors, although the multiorbital electronic correlations also come into play.**

[†]*To whom correspondence should be addressed. Email: clsong07@mail.tsinghua.edu.cn, xucunma@mail.tsinghua.edu.cn, qkxue@mail.tsinghua.edu.cn*

Trivalent fullerides $A_3C_{60}$ ($A$ = alkali metals) have historically been, albeit not universally[10,11], thought of as conventional Bardeen-Cooper-Schrieffer (BCS) superconductors[12,13]. However, the recently revealed dome-shaped dependence of $T_c$ on interfullerene separation[2-5], in analogy with unconventional high-$T_c$ superconductors from cuprates to ferropnictides, shows the importance of electronic correlations in the expanded $Cs_3C_{60}$ polymorphs[8,9]. Alkali fullerides, in which a conventional phonon-mediated superconductivity encounters with Mott physics, revive hope for understanding the pairing mechanism and extraordinary electronic phases (e.g. pseudogap) in strongly correlated high-$T_c$ superconductors. With air sensitivity and phase separation of $A_3C_{60}$, experiments on fulleride superconductors have been largely confined to magnetization and NMR measurements[2-6], which provide very limited insights into the superconducting state. The microscopic mechanism of fulleride superconductivity with the conventional phonon-mediated pairing[12,13] or unconventional electronic pairing[2-4,8,10,11], or a synergy between them[5,9,14], remains controversial.

Unlike quasi-2D character of cuprates and ferropnictides, the fullerides exhibit a unique SMIT in 3D materials[2-5]. An interplay between strong electronic correlations and molecular Jahn-Taller (JT) instability may reduce the dimensionality of low-lying states from 3D to 2D[15-17]. This matches with the high upper critical of 90 teslas measured near the SMIT in $Rb_xCs_{3-x}C_{60}$[18]. A fundamental question thus naturally arises as to how the reduced dimensionality affects superconductivity in fullerides[19], or whether the superconductivity can survive down to the 2D limit as cuprates[7].

Addressing the above questions demands both well-controlled fulleride samples and diverse measurement techniques. Using state-of-the-art molecular beam epitaxy (MBE) technique, we prepared single crystalline films of face-centered cubic (fcc) $K_xC_{60}$ ($x \sim 3$) on graphitized SiC(0001) substrates (Supplementary Section 1 and 2), schematically shown in Fig. 1a,b. The epitaxial films allow themselves being precisely tunable in electronic correlations (characterized by Hubbard $U$) by varying the layer index and electron doping $x$, and being studied by *in-situ* STM. Figure 1c typifies the STM topography of trilayer $K_3C_{60}$ that is $C_{60}$-terminated and unreconstructed (111)-1 × 1 surface (white rhombus). In marked contrast with bulk fcc fullerides[2,4], the tri-star-like $C_{60}$ molecules (with one hexagon pointing up) have the same orientation that indicates no orientational (merohedral) disorder in epitaxial $K_3C_{60}$ films[20]. The closely-packed $C_{60}$ molecules are spaced 10.0 ± 0.1 Å apart, close to the reported value of ~ 10.07 Å for bulk $K_3C_{60}$[12].

Figure 1d shows the spatial dependence of the tunneling conductance spectrum $\sigma(E = eV) \equiv$ d$I$/d$V$ in trilayer K$_3$C$_{60}$ at 2.5 K ($E$ is the electron energy), which reflects the quasiparticle density of states (DOS) and measures the superconducting gap at Fermi energy ($E_F$). Despite somewhat heterogeneity in the coherence peaks, the quasiparticle DOS for d$I$/d$V$ spectra are completely vanishing over an extended energy range $|E| \leq 4.5$ meV, indicating a fully gapped superconductivity. At some locations, the superconducting gap exhibits pronounced coherence peaks and could be fairly well fitted with a single BCS-type isotropic $s$-wave gap function (Fig. 1e)[21]. The mean gap magnitude Δ, half distance between the two coherence peaks, is estimated to be 5.7 ± 0.3 meV. Figure 1f plots the temperature dependence of d$I$/d$V$ spectra from 2.5 K to 40 K which behaves very unusually compared with conventional superconductors, although the superconducting gap and coherence peaks are suppressed at elevated temperatures, as anticipated. A corresponding analysis of the normalized tunneling d$I$/d$V$ spectra as a function of temperature in Fig. 1g reveals an anomaly in gap depth (blue squares) around 22 K that we refer to as the superconducting transition temperature $T_c$, at which the coherence peaks coincidentally disappear (red circles). Across $T_c$, the superconducting gap evolves continuously into a normal state quasiparticle gap with an energy scale that changes little with temperature. This is reminiscent of the pseudogap phenomena in the underdoped cuprates[22] that just become smeared out with increasing temperature above $T_c$.

Only a triple layer K$_3$C$_{60}$ sustains an $s$-wave superconductivity with both $T_c$ and Δ exceeding those ($T_c \sim 19$ K, Δ = 4.4 meV) of its bulk counterparts[21,23], as well as coexistence with a cuprate-like pseudogap within the microscopic scale, is unexpected and has been unambiguously corroborated by the observation of vortices under magnetic field $B$ (Fig. 2a). By fitting the radial dependence of the normalized zero-energy conductance $\sigma_{0,N}$ in the vicinity of vortices (Supplementary Section 3), the superconducting coherence length ξ is deduced and azimuthal angle θ-independent (Fig. 2b,c). This is consistent with isotropic $s$-wave symmetry in K$_3$C$_{60}$, since ξ ~ 1/Δ in BCS theory. The mean ξ = 2.6 ± 0.5 nm matches fairly well with ξ ~ 2.6 nm from magnetization measurements of bulk K$_3$C$_{60}$[24]. Figure 2d displays a series of d$I$/d$V$ spectra straddling an individual vortex at 2 T. Vortices suppress the coherence peaks and induce no bound state inside the vortex cores, but with a remaining pseudogap (red line). We emphasize that such pseudogap exists within all vortices we investigated at various $B$ (Supplementary Section 3), and must be inherent to trilayer K$_3$C$_{60}$. In marked contrast to cuprates[22], no spatially modulated electronic charge density is observed in trilayer $K_3C_{60}$ and the bulk counterpart[25], which highlights the cause of pseudogap in high-$T_c$ superconductors to be opposed to a charge density wave (CDW) cause.

To shed light on the enhanced superconductivity and emergent pseudogap in trilayer $K_3C_{60}$, we investigate layer-dependent structural and electronic properties of K-doped fullerides. Unlike the trilayer one, both monolayer and bilayer $K_3C_{60}$ display a $\sqrt{3} \times \sqrt{3}$ superstructure (Fig. 3a,b). Figure 3c elaborates spatially-averaged d$I$/d$V$ spectra at varied $K_3C_{60}$ layer index. Note that the 9 ML-thick $K_3C_{60}$ exhibits metal-like electronic DOS with two sharp peaks below and above $E_F$, which follows the calculated DOS profile for fully ordered $A_3C_{60}$[26]. This not only hints merohedral disorder-free superconductor $K_3C_{60}$, as observed, but also shows that the band structure of 9 ML $K_3C_{60}$ is close to that of bulk $K_3C_{60}$. At reduced layer index, a DOS dip or insulating energy gap is noticeable near $E_F$ and increases in size (Fig. 3c), hallmarks of a metal-insulator transition. The observed gaps of a few hundreds of meV are too large to be ascribed as possible CDW associated with the $\sqrt{3} \times \sqrt{3}$ superstructures in monolayer and bilayer $K_3C_{60}$ (Fig. 3a,b). Instead we propose that the gap arises from the significantly enhanced electronic correlations due to the poor screening of $K_3C_{60}$ at the 2D limit (Fig. 3d). A strong Coulomb repulsion $U$, assisted by the JT effects, splits the $t_{1u}$-derived band of $C_{60}$ into subbands, and the half-filled $K_3C_{60}$ becomes a Mott JT insulator (MJTI) with charge gap opening near $E_F$, as the $Cs_3C_{60}$ behaves at ambient pressure[3-5]. By measuring $U$ between the upper Hubbard bands (UHB, marked by red triangles) and lower Hubbard bands (LHB, marked by blue triangles) in Fig. 3c and band width $W$ (Supplementary Section 4), one sees that $U/W$ increases with reducing $K_3C_{60}$ layer (Fig. 3d), giving rise to a layer-controlled SMIT ($U/W \sim 1$) between trilayer and bilayer $K_3C_{60}$. The trilayer $K_3C_{60}$ exists on the verge of SMIT and a small enhancement of $U$ renders the bilayer $K_3C_{60}$ non-superconducting (Supplementary Section 5).

A similar SMIT has previously been identified in the expanded $Cs_3C_{60}$ and $Rb_xCs_{3-x}C_{60}$ ($0.35 \leq x < 2$) trivalent fullerides through a continuous tunability of the bandwidth $W$, associated with the interfullerene separation[2-5]. Alternatively, the SMIT realized here is predominantly governed by dimension-controlled $U$ in $K_3C_{60}$ films (Fig. 3d). Figure 3e shows the temperature-dependent d$I$/d$V$ spectra on 9 ML $K_3C_{60}$. At cryogenic temperatures, the overall spectral shape again features an $s$-wave superconducting gap, whereas $\Delta = 4.8 \pm 0.2$ meV and $T_c \sim 18.4$ K (Fig. 3f) are smaller than those of trilayer $K_3C_{60}$. More remarkably, the pseudogap, albeit faint, remains above $T_c$ and within vortices (Fig. 3e,g). Figure 3h compares the pseudogap states between 3 ML and 9 ML $K_3C_{60}$. The pseudogap is suppressed in thick $K_3C_{60}$, which, together with its nonobservability in bulk fullerides[21,25], evidently indicates that the pseudogap could be a general phenomenology of 2D superconductors[27].

In what follows, we examine the superconductivity of $K_xC_{60}$ as the electron doping deviates from half-filling. Tetrahedral K vacancies emerge as dark windmills as $x < 3$ (Fig. 4a), whereas excess K dopants appear and occupy the octahedral sites as $x > 3$ (Fig. 4b) (Supplementary Section 6 gives the detailed structural analysis). Evidently, the $s$-wave superconducting gap is little disturbed by both impurities (Fig. 4c), and no in-gap state is found (blue and red curves). Such observations indicate the robust superconductivity against impurities, and strongly support the conventional phonon-mediated electron pairing in trilayer $K_3C_{60}$, although it exists on the verge of SMIT.

It is also interesting that K vacancies suppress superconductivity more significantly than excess K dopants when the d$I$/d$V$ spectra on defect-free regions of $K_xC_{60}$ (dashed lines) are compared with that of stoichiometric $K_3C_{60}$ (black line) in Fig. 4c. To confirm this finding and discuss its possible cause, we tuned $K_xC_{60}$ stoichiometry (layer index) that enables to systematically track the variations of Δ and $T_c$ over a wide range of electron doping $x$ from 2.5 to 3.6 (electronic correlations $U$) (Supplementary Section 4 and 6). In Fig. 5, we plot Δ (circles) and $T_c$ (squares) *versus* $x$ electronic phase diagram, and discover that in the superconducting state Δ scales with $T_c$ (lower panel) and exhibits dome-shaped variations with peaked values at half-filling. A minor but significant distinction between 3 ML and 9 ML fullerides is that the superconductivity in trilayer $K_xC_{60}$ films decreases extremely abruptly below half-filling. Above $x \sim 3.6$, an insulating phase characteristic of tetravalent fullerides emerges and becomes dominant at $x = 4$ (Supplementary Section 6).

The unusual phase diagram reveals important aspects of superconductivity in fulleride solids. First, the superconductivity declines smoothly as the electron doping $x$ deviates from half-filling in bulk-like 9 ML $K_xC_{60}$. This finding, not necessarily contradictory to the BCS-Eliashberg theory of phonon-mediated superconductivity, contrasts sharply with an early study[28], where a conclusive dependence of $T_c$ vs doping was frustrated by the sample diversity with and without merohedral disorders. However, we find it nontrivial to understand the superconducting domes solely from $x$-dependent DOS variation at $E_F$[13], since the half-filled electronic DOS of $K_3C_{60}$ shows a shoulder and even minimum at $E_F$ (Fig. 3c). Therefore, the DOS($E_F$) and superconductivity would be enhanced as $x > 3$, at odds with what we found. We therefore speculate that if the BCS-Eliashberg theory is applicable to fulleride superconductors, either phonon spectrum or electron-phonon coupling or both should vary substantially with doping that merits further experimental investigation.

Secondly, we believe it is unlikely that the gap shrinkage is due to some form of disorder effects away from half-filling, as the $\Delta \sim x$ evolution appears essentially layer-dependent and of great asymmetry relative to $x = 3$ in trilayer fullerides. Another possible explanation is that the dome-shaped superconductivity might be associated with $x$-dependent electronic correlations[9]. Given the fact that the electronic DOS and $U$ crucially rely on $K_xC_{60}$ layer index (Supplementary Section 4), there exists little opportunity that $\Delta$ coincidently reaches its maximum at half-filling for both 3 ML and 9 ML fullerides. Meanwhile, the doping-dependent electronic phase diagram unambiguously rules out the possibility that fulleride superconductivity results from an accidental doping of $A_3C_{60}$ MJTI[10].

A view of fulleride superconductors that has received increasing attention is that the electron-phonon coupling and multiorbital electronic correlations together bring about the high-$T_c$ superconductivity[5,9,14]. Within this and related pictures, it has been theoretically revealed that away from half-filling $T_c$ (or $\Delta$) drops slowly for small $U/W$ unless $x \sim 2$ or 4 is approached, which agrees qualitatively with our findings in 9 ML $K_xC_{60}$. However, this model fails to catch the dome asymmetry and steep $\Delta(x)$ variation below half-filling in trilayer $K_xC_{60}$ (a reduction of $x$ by $\sim 0.1$ wholly kills superconductivity), due to its oversimplification. In reality, the Hubbard $U$ reduces with $x$ (Supplementary Section 4). Below half-filling, a small increase in electronic correlations would localize electrons and significantly suppress superconductivity in strongly correlated 3 ML fullerides, leading to the dome asymmetry. Nevertheless, the $s$-wave symmetry retains over the entire phase diagram (Supplementary Fig. 8), demonstrating a phonon-mediated electron pairing in fullerides irrespective of $U$. A primary role of electronic correlations may be that they increase the quasiparticle mass and decrease $W$. The reduction of $W$ increases the electronic DOS at $E_F$, which is most probably the cause of enhanced superconductivity in trilayer $K_3C_{60}$.

As a final remark, we comment on the reduced gap $2\Delta/k_BT_c > 6$ (Fig. 5, lower panel) that exceeds its canonical BCS value of 3.53 and that (5.3) of bulk $K_3C_{60}$ in tunneling experiments[21], indicating extremely strong coupling superconductivity in fulleride films. This matches nicely with their enhanced electronic correlations, since $2\Delta/k_BT_c$ was found to rise abruptly close to the Mott transition in fulleride superconductors[5,29,30]. It thus becomes an important future issue to study theoretically how strong correlation is reconcilable to robust *s*-wave symmetry over the entire phase diagram of fulleride superconductors. Another interesting experimental challenge is to access tunable high-$T_c$ superconductivity down to the monolayer limit by increasing *W* (i.e. using the smaller Na or Li as dopants) or reducing *U* (see the superconductivity of overdoped bilayer $K_{3.37}C_{60}$ in Supplementary Section 5). In any case, our study opens a new avenue to engineer the electronic correlations and create novel electronic states in fullerides. In contrast to fcc $A_3C_{60}$ bulk crystals, the stoichiometric $K_3C_{60}$ film is merohedrally ordered, suffering from no structural and chemical complexity of the cuprates and oxyarsenides. This appears important as it offers an ideal material for understanding superconductivity in strongly correlated electron systems that is conclusively of *s*-wave symmetry in fulleride superconductors.

**Acknowledgements**

This work was financially supported by the Ministry of Science and Technology of China (2017YFA0304600, 2018YFA0305603, 2016YFA0301004), the Natural Science Foundation of China (grant no. 11634007, 11427903, 11774192), and in part by the Beijing Advanced Innovation Center for Future Chip (ICFC).

**Author contributions**

C.L.S., X.C.M. and Q.K.X. conceived the experiments. M.Q.R., S.H. and S.Z.W. performed sample growth and STM experiments. J.Q.F. carried out the substrate preparation. M.Q.R., C.L.S., S.H. and X.C.M. analyzed the data. C.L.S. and M.Q.R. wrote the paper with input from X.C.M. and Q.K.X.. All authors discussed the results and commented on the manuscript.

**Competing financial interests**

The authors declare no competing interests.

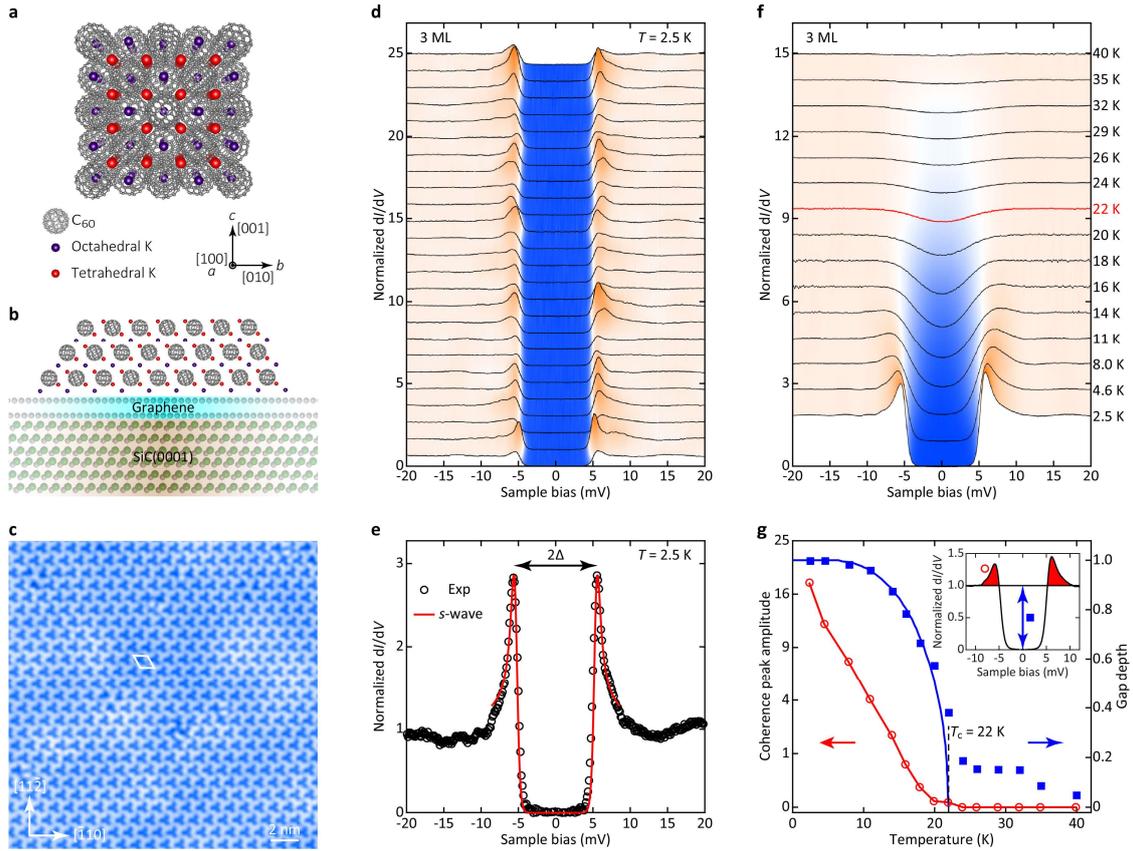

**Figure 1 | STM characterization of trilayer $K_3C_{60}$. a**, Crystal structure of fcc $K_3C_{60}$. Purple and red spheres denote K dopants at the octahedral and tetrahedral sites, respectively. **b**, Schematic (side view) of $K_3C_{60}$(111) epitaxial films on graphitized SiC(0001) substrate. **c**, STM topography of trilayer $K_3C_{60}$(111), acquired at sample bias $V$ = 1.0 V and a constant tunneling current of $I$ = 10 pA. **d**, Grid d$I$/d$V$ spectra in a field of view of 20 nm × 20 nm, revealing uniformity of full-gap superconductivity. The set point is stabilized at $V$ = 20 mV and $I$ = 200 pA, unless other noted. For comparison, each d$I$/d$V$ spectrum is normalized to an extracted cubic background by fitting the raw conductance beyond the superconducting gap ($|V|$ > 10 mV). This normalization procedure is used throughout. **e**, d$I$/d$V$ spectrum at 2.5 K and its best fit (red curve) to a single isotropic $s$-wave superconducting gap with Δ = 5.4 meV. **f**, Spatially-averaged tunneling d$I$/d$V$ spectra as a function of temperature, presenting a marked pseudogap above $T_c$ = 22 K (see the red curve). **g,** Measured dependence on temperature of coherence peak amplitude (circles) and gap depth (squares), defined as the integration over red-shaded regions and difference between unity and normalized zero-energy conductance $\sigma_{0,N}$ (see inset), respectively. At elevated temperatures, the coherence peak amplitude reduces to zero at $T_c$ = 22 K, above which the pseudogap gives rise to a finite gap depth up to 40 K.

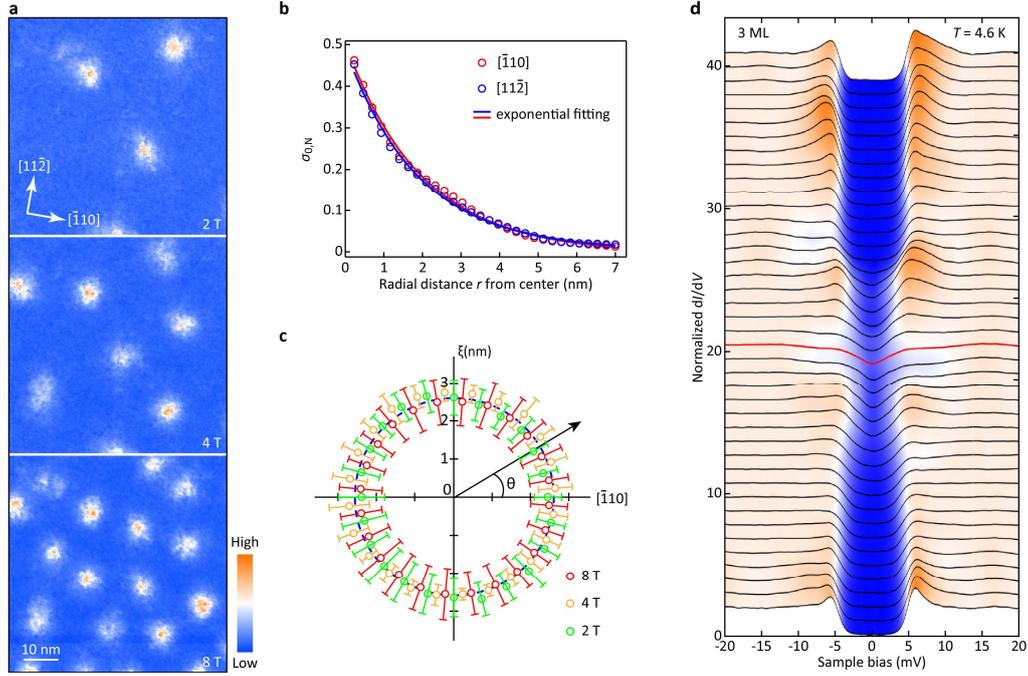

**Figure 2 | Magnetic vortex states. a**, Spatial maps (300 pixels × 300 pixels) of zero-energy conductance $\sigma_0(V = 0)$, revealing the vortices (i.e. the orange/white regions with enhanced $\sigma_0$) at $B$ = 2 T (top), 4 T (middle) and 8 T (bottom) in the same field of view of 63 nm × 63 nm. **b**, Vortex-induced $\sigma_{0,N}$ versus the radial distance $r$ from the center of a vortex at 8 T. Solid lines are the best exponential fits ($\sigma_{0,N} \sim \exp(-r/\xi)$) and give the superconducting coherence length of $\xi$. **c**, Azimuthal dependence of $\xi(\theta)$. The statistical errors of $\xi$ indicate the standard derivations of $\xi(\theta)$ obtained for different vortices at a given magnetic field. **d**, Conductance spectra measured at $T$ = 4.6 K and $B$ = 2 T along a 20-nm trajectory across a vortex core in trilayer $K_3C_{60}$. The thick red line indicates the pseudogap at the vortex center.

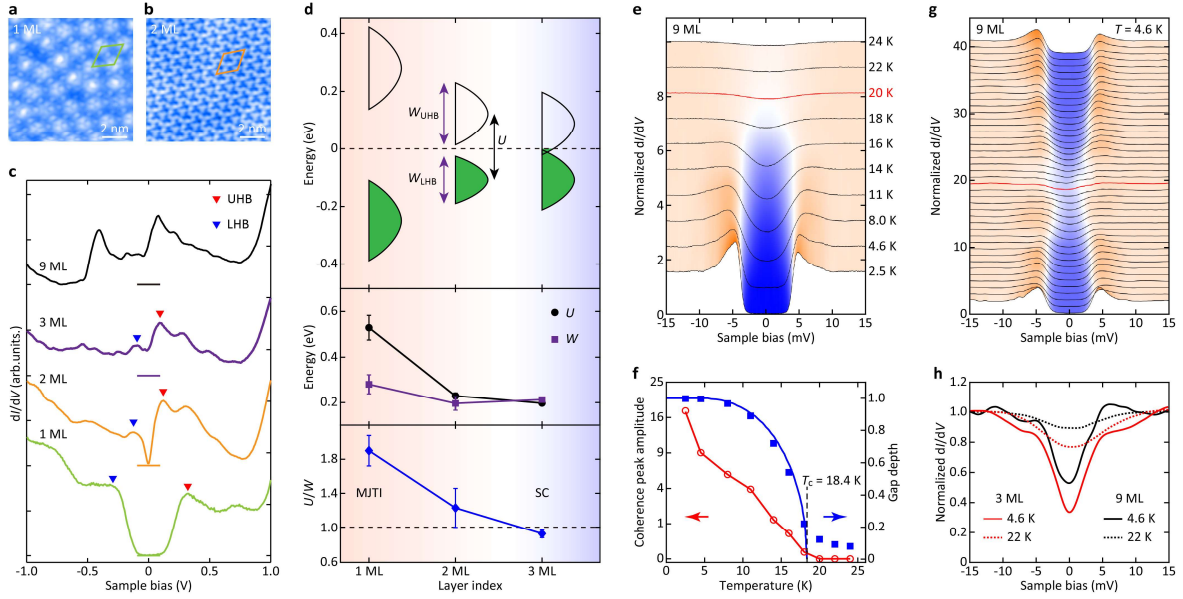

**Figure 3 | SMIT at the 2D limit. a,b**, STM topographies of monolayer and bilayer $K_3C_{60}$ ($V$ = 1.0 V, $I$ = 10 pA) exhibiting a $\sqrt{3} \times \sqrt{3}$ superstructure (marked by the rhombuses). **c**, Layer-dependent tunneling d$I$/d$V$ spectra over a wide energy range of ± 1.0 eV. Setpoint: $V$ = 1.0 V and $I$ = 100 pA. At the 2D limit, the enhanced electronic correlations induce a DOS dip and even insulating energy gap around $E_F$, characterized by the Hubbard $U$ between the UHB (red triangles) and LHB (blue triangles). **d**, Schematic energy bands (top panel) with only the UHB (unfilled) and LHB (green) shown, measured Hubbard $U$, $W$ (middle panel) and $U/W$ (lower panel) of $K_3C_{60}$ as a function of layer index. The statistical errors indicate the standard derivations of $U$ and $W$ measured in various regions. **e,f**, Temperature-dependent tunneling d$I$/d$V$ spectra of 9 ML $K_3C_{60}$, revealing a mild pseudogap (red curve) above $T_c$ = 18.4 K. **g**, Conductance spectra measured at $T$ = 4.6 K and $B$ = 2 T along a 20-nm trajectory across a vortex core in 9 ML $K_3C_{60}$ film. The red line marks the d$I$/d$V$ curve at the vortex center. **h**, Comparison of pseudogaps between 3 ML and 9 ML $K_3C_{60}$ at 4.6 K (acquired within the vortices) and 22 K, respectively.

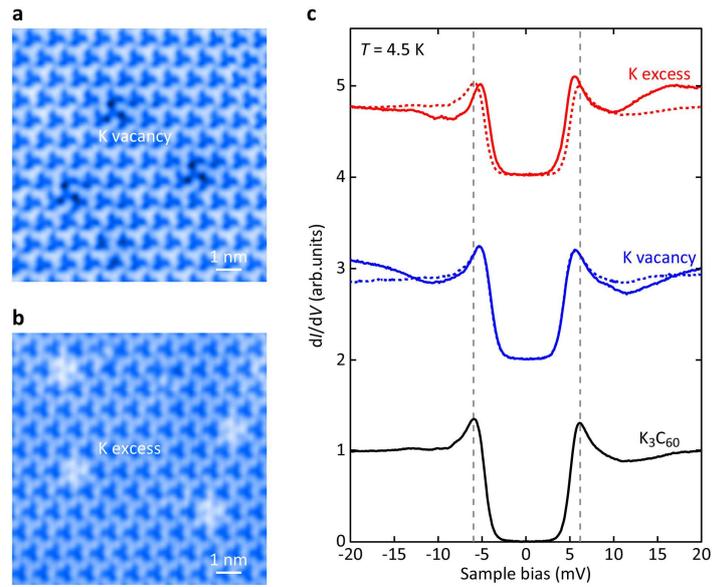

**Figure 4 | Superconductivity away from half-filling. a,b**, STM topographies ($V$ = 1.0 V, $I$ = 10 pA) of trilayer $K_xC_{60}$ with tetrahedral K vacancies ($x$ < 3) and octahedral excess K dopants ($x$ > 3), respectively. **c**, Site-resolved tunneling d$I$/d$V$ spectra on K defects (blue and red lines), defect-free regions of trilayer $K_xC_{60}$ (dashed lines) and stoichiometric $K_3C_{60}$ (black line) at 4.6 K. For every impurity type, at least five impurities have been studied that consistently reveal no in-gap states. The vertical dashed lines are guides to the eye.

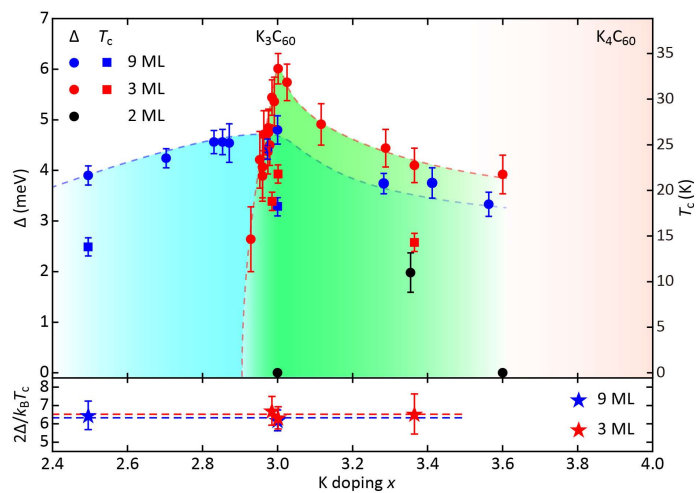

**Figure 5 | Phase diagram.** Electronic phase diagram of fcc $K_xC_{60}$ showing the evolution of $T_c$ (squares), superconducting gap Δ (solid circles) and reduced gap $2Δ/k_BT_c$ (stars, lower panel) at 4.6 K, as a function of K doping $x$. Note that the Δ and thus $2Δ/k_BT_c$ are overestimated by ∼ 6% due to the thermal broadening effect at 4.6 K. The colored symbols distinctively mark the $K_xC_{60}$ fullerides with varied thickness (2 ML: black; 3 ML: red; 9 ML: blue). Gradient shading from green (cyan) to white schematically draws the superconducting dome of 3 ML (9 ML) $K_xC_{60}$ with a peaked $T_c$ and Δ at $x$ = 3. Dashed lines correspond to the experimental tracks for Δ and $2Δ/k_BT_c$ (3 ML: red; 9 ML: blue). The calculated $T_c$ has an uncertainty of 1.0 K and the statistical errors in $x$ (< 0.5%) are smaller than the symbol size.

*Supplementary Materials for*

**Coexistence of full-gap superconductivity and pseudogap in two-dimensional fullerides**

*Ming-Qiang Ren, Sha Han, Shu-Ze Wang, Jia-Qi Fan, Can-Li Song, Xu-Cun Ma, and Qi-Kun Xue*

**CONTENTS**

### I. MATERIALS AND METHODS

All experiments were performed in an ultrahigh vacuum cryogenic (down to 2.5 K) scanning tunneling microscopy (STM) integrated with a superconducting magnet perpendicular to sample surface ($B$ is up to 8 T), and a molecular beam epitaxy (MBE) chamber for *in-situ* sample preparation. Both STM and MBE have a base pressure better than $3.0 \times 10^{-10}$ Torr. $C_{60}$ molecules were evaporated from a standard Knudsen diffusion cell and grew layer-by-layer on nitrogen-doped SiC(0001) wafers (0.1 Ω·cm) at 473 K, which were pre-graphitized by thermal heating (up to 1600 K) to form a bilayer graphene-terminated surface. Potassium (K) atoms were then deposited on the $C_{60}$ epitaxial films at a low temperature of ∼ 200 K step by step, followed by > 3 hours of post-annealing under 300 K. The layer index $n$ of $K_xC_{60}$ multilayers ($n ≤ 3$) is determined from the step height, STM topographies and d$I$/d$V$ spectra, which varies significantly with $n$. The flux rate of $C_{60}$ is thus calculated by dividing the coverage of fulleride films by growth time, and is used as a reference for determination of the nominal thickness for thicker fullerides (e.g. $n = 9$ in the main text) throughout the experiments.

In K-doped fullerides, the electron doping $x$ (2.4 < $x$ < 3.4) is calculated directly from the areal density of K vacancies (Fig. 4a in the main text) or excess K dopants (Fig. 4b in the main text). For $x$ ∼ 3, there exist little defect that leads to the merohedrally ordered trivalent fulleride films (Fig. 1c). Here, a single K vacancy (excess) is reasonably considered as one missing (additional) K dopant relative to $K_3C_{60}$. For higher doping, $x$ is estimated from the coverage of K clusters, since the excess K dopants are individually undistinguishable. By this method, the estimated $x$ has a statistical error of < 0.5%.

To accurately characterize the superconductivity and electronic structure of $K_xC_{60}$ films, special measures such as grounding and shielding were taken to optimize the stability and spectroscopic resolution of our STM facility. Polycrystalline PtIr tips were used after careful calibration on Ag films grown on Si(111). All STM topographic images were taken in a constant current mode. Tunneling d$I$/d$V$ spectra and electronic DOS maps were acquired using a standard lock-in technique with modulation frequency $f$ = 975 Hz, while the modulation amplitudes were 0.2 meV and 20 meV for measuring the superconducting gaps and wider-energy-range (± 1.0 eV) d$I$/d$V$ spectra, respectively.

## II. CRYSTAL STRUCTURE OF EPITAXIAL $K_3C_{60}$ FILMS

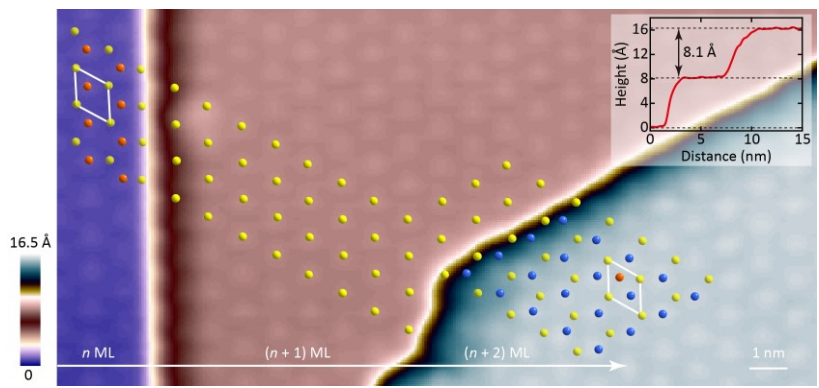

**Supplementary Figure 1 | Stacking sequence and fcc crystal structure of epitaxial $K_3C_{60}$ films.** STM topography ($V$ = 3.0 V, $I$ = 10 pA) showing three consecutive $K_3C_{60}$ layers. From left to right, the layer

index is defined as *n*, (*n* + 1) and (*n* + 2), with their respective molecular centers marked by orange, yellow and blue spheres. Note that the $C_{60}$ molecules in *n* ML and (*n* + 2) ML occupy two symmetry-inequivalent sites in the unit cell of the middle layer (white rhombus), revealing a (111)-orientated fcc crystal structure. From the inserted profile along the white arrow, the inter-layer spacing is measured to be 8.1 ± 0.1 Å, as expected for the fcc-structured $K_3C_{60}$ with a lattice constant of ∼ 14.2 Å[17].

## III. PSEUDOGAP WITHIN VORTICES

To quantify vortex-induced states, the zero-energy conductance $\sigma_0(V = 0)$ maps (Fig. 2a in the main text) are normalized to the normal-state ones (i.e. $\sigma(V = \pm 12 \text{ meV})$), and we therefore obtain the normalized $\sigma_{0,N}$ as

$$\sigma_{0,N} = \frac{\sigma_0(V=0)}{(\sigma(V=12 \text{ meV})+\sigma(V=-12 \text{ meV}))/2}.$$

Here $\sigma_{0,N}$ will be increased to 1 within vortices if no vortex bound state is formed there (dirty-limit superconductivity). Otherwise, as the pseudogap states develop within vortices just like cuprates[22], $\sigma_{0,N}$ is always less than unity and can never reach 1 there. This happens to occur in Supplementary Fig. 2, indicative of ubiquitous pseudogap state in $K_xC_{60}$ films.

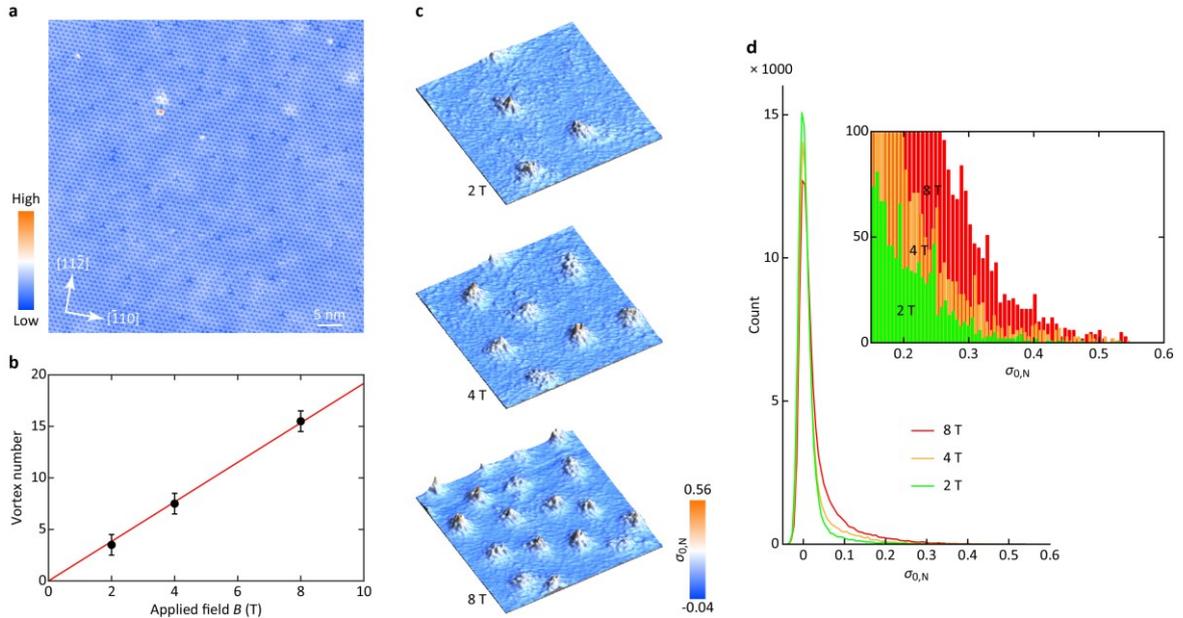

**Supplementary Figure S2 | Ubiquitous pseudogap state within vortices. a**, STM topographic image (*V* = 1.0 V, *I* = 10 pA) of trilayer $K_xC_{60}$ (*x* ∼ 2.976), over which the magnetic vortices are imaged. **b**, Measured vortex number versus *B*, matching with the expectation (red line) for a field of view

of 63 nm × 63 nm in **a**. **c**. 3D representation of $\sigma_{0,N}$ maps at varied $B$. **d**. Appearance counts versus $\sigma_{0,N}$ in **c**. The emergence of magnetic vortices leads to a peak asymmetry at $\sigma_{0,N} = 0$ that gets more prominent at elevated $B$. Inset shows a magnified histogram of $\sigma_{0,N} \ll 1$ around vortices, hallmarks of the emergence of pseudogaps within every vortex.

## IV. FILLING DEPENDENCE OF ELECTRONIC DOS

### Electronic DOS *versus* doping & layer index

A careful examination of doing-dependent d$I$/d$V$ spectra in Supplementary Fig. 3 reveals that the Hubbard $U$ decreases with $x$ in bilayer K$_x$C$_{60}$ due to the enhanced screening from itinerant electron carriers. Followed by the $U$ reduction, a transition from MJTI to metal or superconductor happens between $3 < x < 3.36$, as evidenced by the emergent DOS at $E_F$ (Supplementary Fig. 3a). Actually, a signature of superconductivity is observed in bilayer K$_x$C$_{60}$ with $x = 3.36$ (Supplementary Fig. 6). The doping-induced reduction in $U$, albeit faint, holds true in trilayer K$_x$C$_{60}$. This is further rationalized by the fact that the DOS dip at $E_F$, due to strong electronic correlations, gradually gets shallower with reducing $U$ and almost vanishes as $x > 3.25$ (Supplementary Fig. 3b). Without loss of generality, a similar $E_F$-near DOS dip emerges in 9 ML K$_{2.83}$C$_{60}$ and deepens in K$_{2.703}$C$_{60}$, whereas there exists no gap at and above half-filling (Supplementary Fig. 3c). Taken altogether, a fine control over electronic correlation $U$ via electron filling and layer index has been realized in epitaxial K$_x$C$_{60}$ films. Evidently, the electronic correlations of K$_x$C$_{60}$ become stronger with reducing $x$ and layer index.

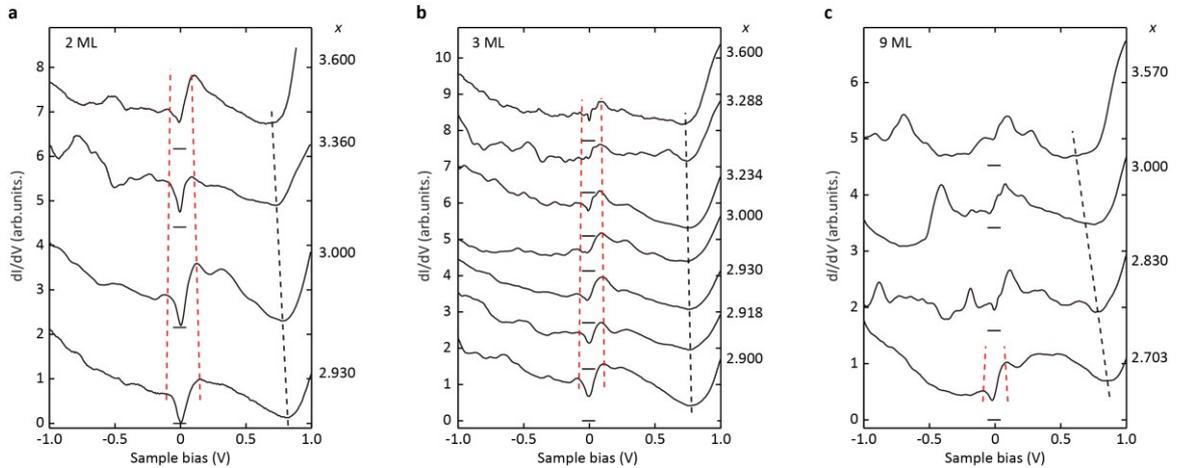

**Supplementary Figure 3 | Electronic DOS. a-c**, Spatially-averaged d$I$/d$V$ spectra over a wide energy range of ± 1.0 eV ($V$ = 1.0 V, $I$ = 200 pA) as a function of electron doping $x$, measured on **a**, 2 ML; **b**,

3 ML; **c**, 9 ML K$_x$C$_{60}$. The red dashed lines mark the doping evolution of UHB and LHB, with the black dashed ones signifying K-increased electron filling (to wit the progressive upward shifting of $E_F$ with increasing *x*).

## Fitting tunneling spectra for Mott parameters

Tunneling spectra in strongly correlated K$_x$C$_{60}$ films reveal Mott-Hubbard-like band structure near $E_F$ and satellite peaks (probably caused by the JT effects) beyond the Hubbard gaps (Fig. 3c in the main text and Supplementary Fig. 3). In monolayer K$_3$C$_{60}$, the Hubbard *U* is huge and one can extract *U*, onset energies of LHB and UHB straightforwardly from the intersections of linear fits to the d*I*/d*V* spectral weights (red and blue lines in Supplementary Fig. 4a)[31]. Here *U* could be reasonably defined as the energy separation between LHB- and UHB-derived kink/peak in electronic DOS (marked by blue and red triangles, respectively), while the bandwidth *W* is calculated as $W = 1/2 (W_{UHB} + W_{LHB})$, with $W_{UHB}$ and $W_{LHB}$ representing the bandwidths of UHB and LHB, respectively. The same definition is used throughout. As the layer index of K$_x$C$_{60}$ films increases to 2 and 3 (Supplementary Fig. 3b,c), the comparable *U* and *W* mix the UHB and LUB to some extent, and we instead fitted the measured electronic DOS near $E_F$ with three Gaussian profiles, as well justified in Supplementary Fig. 4b,c. As thus, the bandwidths $W_{UHB}$ and $W_{LHB}$ are calculated as two times of the full width at half maxima (FWHM) of the blue and red Gaussian peaks, while the *U* represents the separation between the two Gaussian peaks.

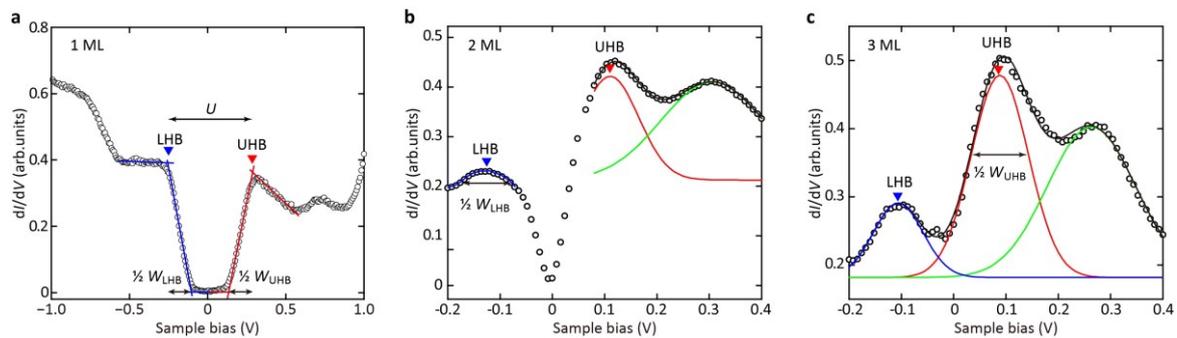

**Supplementary Figure 4 | Fitting tunneling spectra for Hubbard *U* and bandwidth *W*. a-c**, d*I*/d*V* conductance spectra in the vicinity of $E_F$, measured on **a**, 1 ML; **b**, 2 ML; **c**, 3 ML K$_3$C$_{60}$. Black empty circles and lines designate the experimental and fitted curves, respectively. The blue, red and green solid lines denote the decomposed Gaussian functions responsible for the LHB, UHB and satellite peak just above UHB, respectively.

## V. FILLING-CONTROLLED SMIT IN BILAYER FULLERIDES

### Insulating bilayer $K_3C_{60}$

In half-filled bilayer $K_3C_{60}$, the small-energy-scale d$I$/d$V$ spectra are all characterized by an energy gap in the vicinity of $E_F$ (Supplementary Fig. 5a,b). Such a tunneling gap, which slightly varies with the measurement regions, is not always symmetric relative to $E_F$ and exhibits an extremely large gap size of 80 ± 20 meV, as compared to the superconducting gap (Fig. 5 in the main text). Moreover, the gap proves to be not sensitive to an external magnetic field of 8 T and the sample temperature (Supplementary Fig. 5c,g). No signature of vortex is observed in the d$I$/d$V$ maps under a magnetic field of 8 T (Supplementary Fig. 5d-f). Altogether, we conclude that the bilayer fullerides are not superconducting at half-filling due to the enhanced electronic correlations down to the 2D limit.

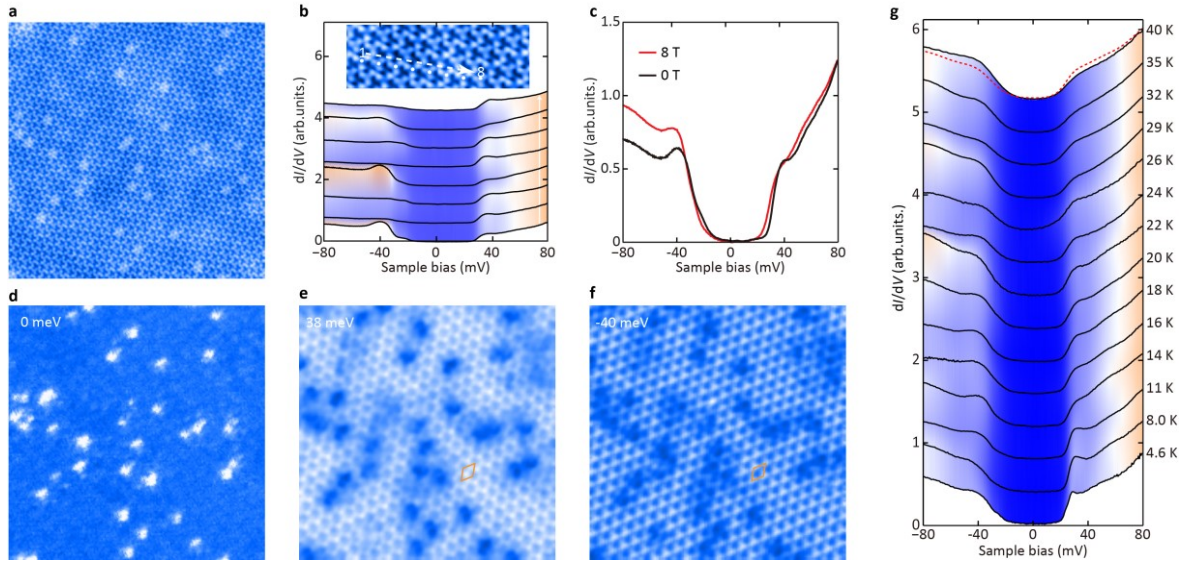

**Supplementary Figure 5 | STM characterization of bilayer $K_3C_{60}$ with half-filling. a**, Topography of bilayer $K_3C_{60}$, interspersed with small amount of excess K dopants ($V$ = 1.0 V, $I$ = 30 pA, 33 nm × 33 nm). **b**, Spatially-resolved tunneling d$I$/d$V$ spectra alone the [$\bar{1}$10] direction (see inset), showing a larger energy gap of ~ 80 meV accompanied with a periodic variation in the spectral intensity near gap edges (i.e. every three $C_{60}$ anions). The periodicity of ~ 30 Å is associated with the √3 × √3 superstructure in bilayer $K_3C_{60}$ (Fig. 3b in the main text). **c**, Averaged d$I$/d$V$ spectra (Set point: $V$ = 100 mV, $I$ = 200 pA, lock-in modulation $\Delta V$ = 1 mV) at $B$ = 0 (black curve) and $B$ = 8 T (red curve). **d-f**, Electronic DOS maps with energy **d**, $E$ = 2 meV, **e**, 38 meV and **f**, -40 meV acquired simultaneously in the same field with **a** at $B$ = 8 T. The tunneling conductance peaks (bright areas in **d**) or valleys

(dark patches in **e**, **f**) match fairly nicely the positions of excess K, assigned as excess K dopant-induced electronic states. Note that a $\sqrt{3} \times \sqrt{3}$ superstructure in electronic DOS is clearly seen in **e** and **f**, which exhibits reversal intensity at occupied and empty states (see the orange rhombuses for $\sqrt{3} \times \sqrt{3}$ superstructure), indicating a possible origin from CDW. **g**, Spatially-averaged d$I$/d$V$ spectra as a function of temperature. The red dashed line shows the convolution of electronic DOS measured at 4.6 K with a Fermi-Dirac function with $T$ = 40 K. Apart from thermal broadening effect, the gap remains robust in shape and size with increasing temperature, indicating a phase transition temperature of $T_s$ > 40 K.

Superconductivity in bilayer $K_xC_{60}$ ($x$ = 3.36)

With increasing $x$ above half-filling, the more itinerant electron carriers enter and reduce the screened Coulomb potential of $K_xC_{60}$, evidenced by a reduction of $U$ and emergent electronic DOS at $E_F$ in Supplementary section 4. Instead of the asymmetric gap in half-filled $K_3C_{60}$ (Supplementary Fig. 5), a bilayer $K_xC_{60}$ film with $x \sim 3.36$ displays a superconducting gap near $E_F$ (Supplementary Fig. 6). Despite somewhat inhomogeneity in coherence peak strength, the superconducting gap exists ubiquitously both on and off K dopants and has a gap size of approximately 2.0 meV. We therefore realize a filling-controlled SMIT in bilayer $K_xC_{60}$ just like that in cuprate superconductors.

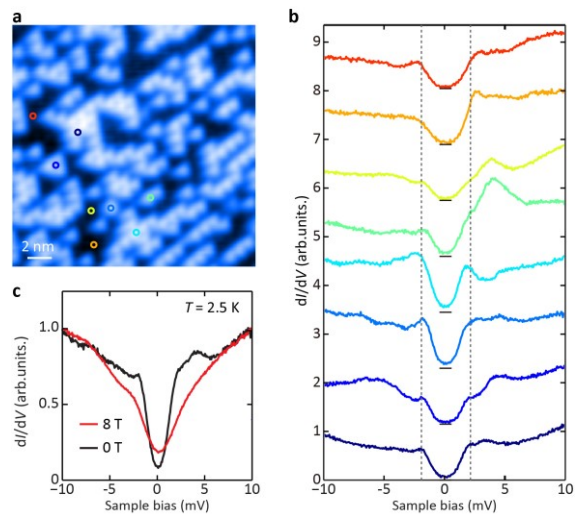

**Supplementary Figure 6 | Evidence of superconductivity in bilayer $K_xC_{60}$ with $x \sim 3.36$. a**, Typical STM topography ($V$ = 3.0 V, $I$ = 30 pA) of bilayer $K_{3.36}C_{60}$, with excess K dopants randomly distributed (bright dots). **b**, A series of d$I$/d$V$ spectra (set point: $V$ = 10 mV, $I$ = 200 pA) measured at $T$ =2.5 K, color-coded to match the probe positions (empty circles) in **a**. The vertical dashed lines mark the

gap edges, whereas the short horizontal bars denote the zero conductance positions for the nearby d$I$/d$V$ spectra. **c**, Spatially-averaged d$I$/d$V$ spectra at $B$ = 0 T (black curve) and $B$ = 8 T (red curve). The tunneling gap is suppressed by an 8 T magnetic field, embodied in the raised DOS around $E_F$ and the killed coherence peaks, and thus rationalized to be associated with superconductivity.

## VI. SUPERCONDUCTIVITY AWAY FROM HALF-FILLING

### Defect registry

In experiment, the excess K dopants occupy a symmetry-equivalent threefold hollow site of the outmost $C_{60}$ layer (Fig. 4b in the main text), for which we conclusively assign them as the octahedral K excesses at the surface (Supplementary Fig. 7a). In contrast, the K vacancies are positioned at the top sites if one ascribes the bright-core-centered tri-star to $C_{60}$ (Registry I in Supplementary Fig. 7b and Fig. 4a in the main text), indicating the registry of K vacancies at the 4$^{th}$ tetrahedral layer (Supplementary Fig. 7a). This is hard to be understood, and we thus propose another possibility of Registry II in Supplementary Fig. 7b. In this scenario, the tri-stars exhibit dim centers just as the un-doped $C_{60}$[31]. The K vacancy occupies another threefold hollow site of the outmost $C_{60}$ layer (the 1$^{st}$ tetrahedral layer) and behaves dark as observed. This appears more reasonable and matches better with defect imaging in the heavily underdoped $K_xC_{60}$ films with smaller $x$ (Supplementary Fig. 7c,d).

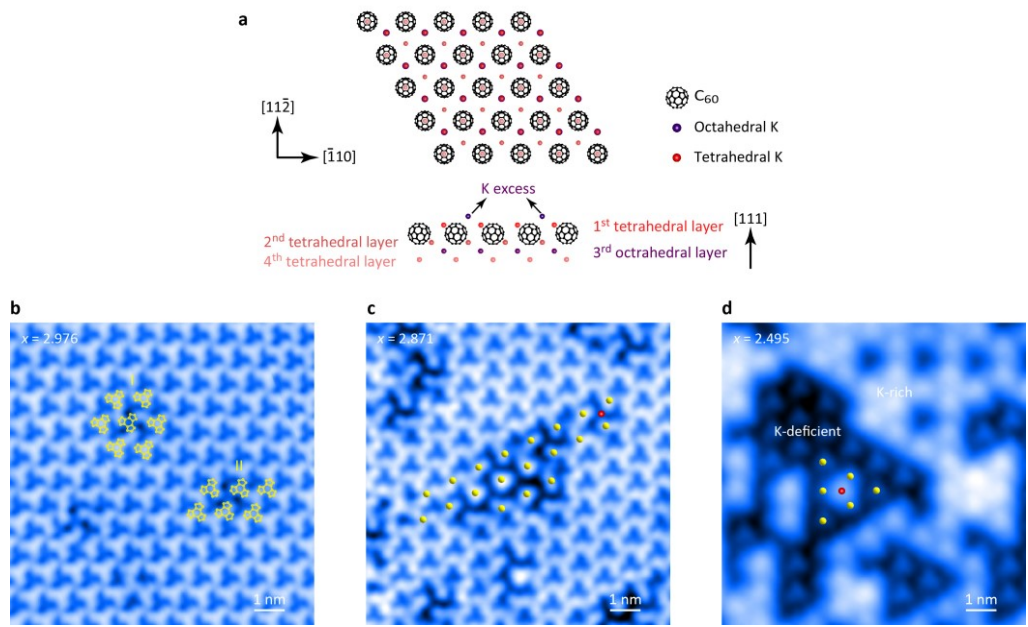

**Supplementary Figure 7 | Registry of K defects away from half-filling. a**, Top and side views of the

(111)-orientated $K_3C_{60}$ fulleride. For simplification only the outermost $C_{60}$ molecular layer and the nearby K dopant layers (differentiated by color and opacity, see the text annotation) are shown. **b**, STM topography of $K_xC_{60}$ (same as Fig. 4a in the main text) exhibiting three single K vacancies. The up pointing hexagon skeleton (yellow patterns) of $C_{60}$ is overlaid onto the image to illustrate two possible registries (I and II) of K vacancy relative to the outermost $C_{60}$ anions. **c,d**, STM topographic images (**c**, $V$ = 1.0 V, $I$ = 10 pA; **d**, $V$ = 3.0 V, $I$ = 10 pA) of $K_xC_{60}$ with $x$ = 2.871 and $x$ = 2.495, respectively. Yellow and red spheres denote the outermost $C_{60}$ and K dopants in the $1^{st}$ tetrahedral layer. Note that the bright spots in K-rich domains denote the $1^{st}$ tetrahedral K dopants and occupy the threefold hollow site of the outmost $C_{60}$ molecules, which exhibit triangle-like patterns at a sample bias of 3.0 $V$ in the K-deficient domain of **d**.

Superconducting gap *versus* doping *x*

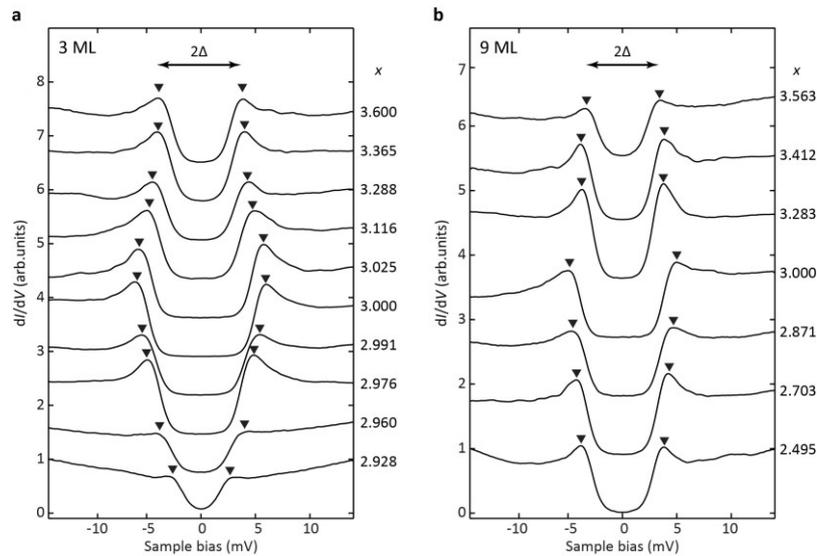

**Supplementary Figure 8 | Evolution of the superconducting gap with K doping *x*. a,b**, Averaged d$I$/d$V$ spectra measured on 3 ML and 9 ML $K_xC_{60}$, respectively. All spectra were measured at $T$ = 4.6 K (set point: $V$ = 20 mV, $I$ = 200 pA). For every $x$, the site-resolved d$I$/d$V$ spectra (>10 curves) are averaged to minimize the effect of spatial inhomogeneity. The coherence peaks are marked by the black triangles for eye guide. The superconducting gap size exhibits dome-shaped variations, with its maximum locked at $x$ = 3 (stoichiometric $K_3C_{60}$), and evolves smoothly with $x$ diverge from 3 for both 3 ML and 9 ML $K_xC_{60}$, as discussed in the main text. Notably, all the d$I$/d$V$ spectra show a fully gapped superconductivity with zero and flat bottoms in d$I$/d$V$ around $E_F$. The parabolic-shaped gap

for *x* = 2.928 trilayer film results from a limited resolution of our system for small Δ and the spatially averaged effect of d*I*/d*V* spectra.

*T*<sub>c</sub> versus doping *x*

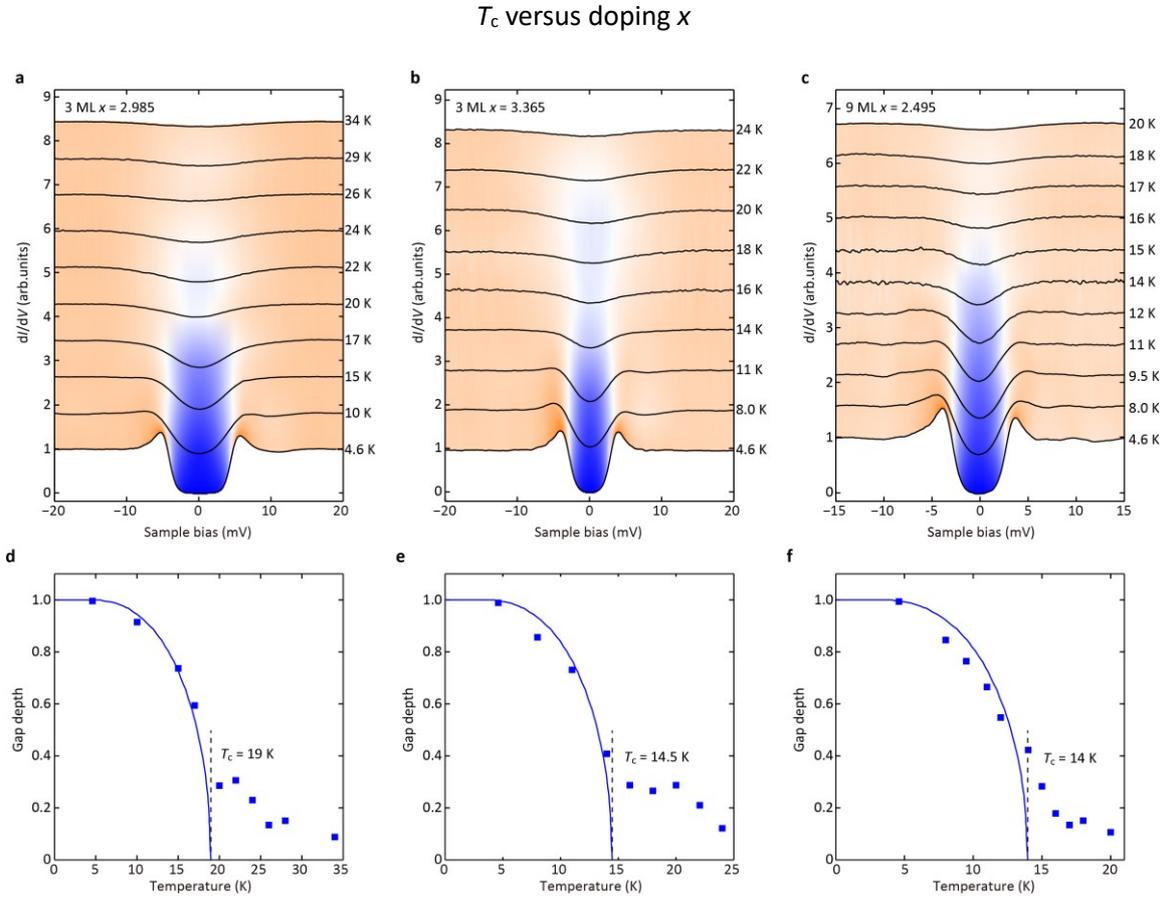

**Supplementary Figure 9 | Temperature dependence of the superconducting gap. a-c**, Spatially-averaged d*I*/d*V* spectra as a function of temperature, measured on **a**, 3 ML, *x* = 2.985; **b**, 3 ML, *x* = 3.365; **c**, 9 ML, *x* = 2.495. All the d*I*/d*V* spectra were normalized to an extracted cubic background, identical to that used in the main text. **d-f**, Measured dependence on temperature of gap depth (blue squares), which behave similarly to the evolution of the reduced superconducting gap $\Delta/\Delta_0$ (*T* = 0) versus *T* in a classic BCS superconductors (blue curves). The best fit gives a *T*<sub>c</sub> = 19 K, 14.5 K and 14 K for the superconducting gaps in **a** - **c**, respectively. Note that the pseudogap exists above *T*<sub>c</sub> and is generic to all the superconducting $K_xC_{60}$ thin films investigated.

Phase separation and insulating $K_4C_{60}$

As the nominal electron doping *x* is increased to critical value of ~ 3.3, the K-doped films separate spatially into the superconducting and insulating domains, as illustrated in Supplementary Fig. 10a.

Unlike tri-star-like molecular pattern in the superconducting domains (Supplementary Fig. 10b), the $C_{60}$ molecules change their orientations with one hexagon-hexagon band pointing up (see the two-lob intramolecular structure) in Supplementary Fig. 10c, hallmarks of tetravalent fullerides[20,31]. Due to JT instability, the triply degenerate $t_{1u}$ orbitals of $C_{60}$ anions split into subbands, thereby leading to charge-disproportionated insulator[32,33]. This agrees well with our observations that the tunneling d$I$/d$V$ spectra reveal a large band gap opening up to 1.2 eV in Supplementary Fig. 10d. Note that $E_F$ in morphology B locates at the boundary of the valence band, signifying $x < 4$[31]. With more K atoms deposited, the superconducting $K_xC_{60}$ gradually decreases in area and evolves into the tetravalent fulleride $K_4C_{60}$ eventually. The region A differs slightly from region B in both morphology and d$I$/d$V$ spectrum, and the reason behind them lies beyond the scope of this study.

For the situation $x < 3$, the deficiency of K dopants was found to drive partial regions of samples into K-vacancy-abundant $K_3C_{60}$, while leaving other regions as the pristine $C_{60}$ (undoped) films. With more K dopants deposited, the $K_3C_{60}$ region grows bigger and the K vacancies decrease in density, until finally evolves into stoichiometric $K_3C_{60}$. The $K_1C_{60}$ and $K_2C_{60}$ compounds don't show up during the entire growth process of multilayer films, indicating that they are less thermodynamically stable than $K_3C_{60}$ on graphene substrate.

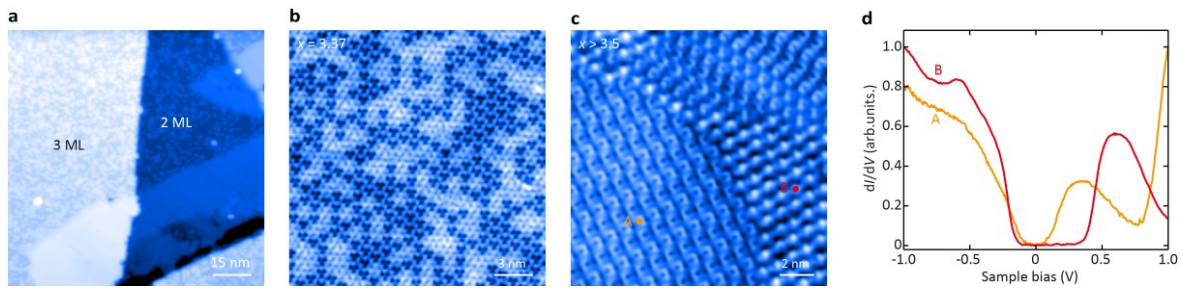

**Supplementary Figure 10 | Phase separation away from half-filling. a**, STM topography ($V$ = 3.0 V, $I$ = 5 pA) exhibiting phase separation between superconducting (rough regions) and insulating (flat regions) domains on both 2 ML and 3 ML $K_xC_{60}$ with $x \sim 3.3$. **b**, STM image ($V$ = 1.0 V, $I$ = 10 pA) in the superconducting domains of trilayer $K_xC_{60}$, $x$ = 3.365. The surface is decorated with random-distributed K dopants and the superconductivity is confirmed by d$I$/d$V$ spectra in Supplementary Fig. 9. **c**, STM image ($V$ = -1.0 V, $I$ = 5 pA) of the insulating 3 ML $K_xC_{60}$, $x > 3.5$, showing two different kinds of surface morphologies as marked by A and B. Note that the tri-star-like $C_{60}$ anions has been replaced by the two-lob intramolecular structure. **d**, Site-resolved d$I$/d$V$ spectra (set point: $V$ = 1.0 V, $I$ = 200 pA) on morphologies A and B, both characterized by an insulating gap.